\documentclass[prd,aps,preprint,amsmath,amssymb,nofootinbib]{revtex4-1}
\usepackage{verbatim,graphics,graphicx,color,slashed,textcomp}
\usepackage[normalem]{ulem} 
\usepackage[colorlinks=true,linkcolor=red,urlcolor=blue,citecolor=blue]{hyperref}

\graphicspath{{figures/}}

\usepackage[utf8]{inputenc}

\newcommand{\GeV}{\mathrm{GeV}}
\newcommand{\TeV}{\mathrm{TeV}}

\begin{document}
	
\title{Weyl Symmetry Inspired Inflation and Dark Matter}
	\preprint{UT-19-07}
	\author{Yong Tang$^{a,b,c,d}$ and Yue-Liang Wu$^{d,e,f,g}$}
	\affiliation{\begin{footnotesize}
		${}^a$School of Astronomy and Space Sciences, University of Chinese Academy of Sciences (UCAS), Beijing, China\\
		${}^b$National Astronomical Observatories, Chinese Academy of Sciences, Beijing, China \\
		${}^c$Department of Physics, University of Tokyo, Bunkyo-ku, Tokyo 113-0033, Japan\\
		${}^d$School of Fundamental Physics and Mathematical Sciences, \\
		Hangzhou Institute for Advanced Study, UCAS, Hangzhou 310024, China \\
		${}^e$International Centre for Theoretical Physics Asia-Pacific, Beijing/Hangzhou, China \\
		${}^f$Institute of Theoretical Physics, Chinese Academy of Sciences, Beijing 100190, China \\
	    ${}^g$School of Physical Sciences, University of Chinese Academy of Sciences, Beijing, China    
\end{footnotesize}}

\begin{abstract}
Motivated by the Weyl scaling gauge symmetry, we present a theoretical framework to explain cosmic inflation and dark matter simultaneously. This symmetry has been resurrected in recent attempts to formulate the gauge theory of gravity. We show the inspired inflation model is well consistent with current observations and will be probed further by future experiments. Furthermore, we clarify and prove the stability of Weyl gauge boson in the general theory with multiple scalars. We show the massive Weyl gauge boson can be a dark matter candidate and give the correct relic abundance.  
\end{abstract}	
	
\maketitle

\section{Introduction}
The accumulated compelling evidence for dark matter (DM) has been challenging the standard model (SM) of fundamental physics for decades. The supporting observations, such as cosmic mircowave background (CMB), large-scale structure, rotation curves, scope from cosmological to galactic scales~\cite{Jungman:1995df, Bertone:2004pz}. For the intrinsic nature of DM, however, we are still lacking sufficient information since the robust evidence is only able to suggest that DM must have gravitational interaction. Nevertheless, explanations of DM would require extensions of SM, either in the sector of particle physics or gravity.  

We also know from experimental measurements that the power spectrum of fluctuations in our universe is almost scale invariant, which indicates Weyl/scaling symmetry may play some role in the theory of inflation that generates the primordial fluctuations. Then it is not unreasonable to expect that Weyl symmetry may be also behind the theory of DM, because symmetry has played a guiding principle for constructing fundamental laws of nature since last century when Weyl first proposed~\cite{Weyl:1919fi} the scaling symmetry and tried to unify the electromagnetic interaction with Einstein's general relativity. The original scale factor has to be modified as a phase to account for the gauge $U(1)$ theory for electromagnetic interaction~\cite{Weyl:1929fm}. $U(1)$ is later generalized to non-abelian theory by Yang and Mills~\cite{Yang:1954ek}, which describes the interactions of all known fundamental particles in SM by incorporating the Higgs mechanism~\cite{Englert:1964et,Higgs:1964pj,Guralnik:1964eu}. Variants of Weyl symmetry, however, still stimulate explorations of theoretical and phenomenological studies, see Refs.~\cite{Zee:1978wi, Adler:1982ri, Fujii:1982ms, Wetterich:1987fm, Cheng:1988zx, Cheng:2004sd, Kaiser:1994vs, Foot:2007iy, Nishino:2009in, Ferrara:2010in, GarciaBellido:2011de, Hur:2011sv,  Kallosh:2013hoa, Bars:2013yba, Farzinnia:2013pga, Holthausen:2013ota, Giudice:2010ka, Kurkov:2013gma, Csaki:2014bua, Iso:2014gka, Guo:2015lxa, Kannike:2015apa, Kannike:2015kda, Salvio:2017xul, Pallis:2018ver, Ferreira:2018qss, Tang:2018mhn, Barnaveli:2018dxo, Ghilencea:2018thl, Kubo:2018kho} for various examples in cosmology and particle physics. Recently, Refs.~\cite{Wu:2015wwa,Wu:2017urh} has shown the original Weyl symmetry can play a crucial role in formulating the gauge theory of gravity. 

In this paper we propose that the original Weyl symmetry can provide a framework to explain the cosmic inflation~\cite{Guth:1980zm, Starobinsky:1980te, Linde:1981mu, Albrecht:1982wi} and DM simultaneously\footnote{Our proposal is different from the scenario where inflaton is identified as dark matter, see Refs.~\cite{Kofman:1997yn, Lerner:2009xg, Mukaida:2013xxa, Khoze:2013uia, Hooper:2018buz, Borah:2018rca, Choi:2019osi, Daido:2017tbr, Daido:2017wwb} for such examples, and is also different from Ref.~\cite{Katsuragawa:2016yir} where DM is identified as the scalaron in $f(R)$ gravity.}. The starting inflationary Lagrangian can be Weyl invariant and responsible for the generation of Planck scale. Theoretical predictions of observables, scalar spectral index and tensor-to-scalar ratio, are consistent with currect experiments and testable in future. After we clarify the stability issue of Weyl gauge boson in the literature~\cite{Cheng:1988zx,Cheng:2004sd,Kashyap:2012kg} and prove in the general framework with multiple scalars, we show the Weyl gauge boson can be identified as a DM candidate, if the coupling is small enough. 

This paper is organized as follows. In Section.~\ref{sec:frame} we first establish the theoretical framework and the relevant notations. Then in Section.~\ref{sec:inf} we illustrate how viable inflation is provided in our formalism. Later in Section.~\ref{sec:dark} we demonstrate the Weyl gauge boson can be a DM candidate and discuss its relic abundance. Finally, we give our conclusions.

Throughout our paper, we use the sign convention for the metric, $\eta_{ab}=(1,-1,-1,-1)$, and natural unit $M_P\equiv 1/\sqrt{8\pi G} =1$. Sometime $M_P$ is written explicitly without confusion.

\section{Framework}\label{sec:frame}
To illustrate the main physical points, we start with the following general Lagrangian with two real scalars, $\varphi$ and $\phi$, and a fermion $\psi$,
\begin{align}\label{eq:act1}
\mathcal{L}\supset \sqrt{-g}\big[& \frac{\alpha}{2} \left(\varphi^{2}R-6\partial_{\mu}\varphi\partial^{\mu}\varphi\right) + \frac{\beta}{2} \left(\phi^{2}R-6\partial_{\mu}\phi\partial^{\mu}\phi\right) + \frac{\zeta_1}{2}D_{\mu}\varphi D^{\mu}\varphi + 
\frac{\zeta_2}{2}D_{\mu}\phi D^{\mu}\phi \nonumber \\
& + \frac{i}{2}\left(\overline{\psi}\gamma^\mu D_\mu \psi -\overline{D_\mu \psi}\gamma^\mu \psi \right) 
+ y\, \varphi \overline{\psi}\psi  + f \,\phi \overline{\psi}\psi -V(\phi, \varphi)-\frac{1}{4g_W^2}F_{\mu\nu}F^{\mu\nu}\big],
\end{align}
where $R$ is the Ricci scalar, the Weyl field $W_\mu\equiv g_W w_\mu$, $g_W$ is the corresponding gauge coupling, $F_{\mu\nu}=\partial_{\mu}W_\nu-\partial_{\nu}W_\mu$ and the {\it covariant derivative} $ D_{\mu}=\partial_{\mu}-W_{\mu}$, $y$ and $f$ are Yukawa couplings. More complete Lagrangian can be found in Refs.~\cite{Wu:2015wwa, Wu:2017urh} where gravity is formulated as a gauge theory of the fundamental field $\chi_{\mu}^{a}$ with its connection to metric, $\chi_{\mu}^{a}\chi_{\nu}^{b}\eta_{ab}=g_{\mu\nu}$. The potential $V$ can have a general form of $\sum_{i=0}^{4}c_{i}\phi^{i}\varphi^{4-i}$. The parameters $\zeta_i$ in the front of scalar kinetic terms can be positive, negative or zero. Note that negative  $\zeta_i$ is not necessary associated with theoretical issues, as long as the total energy of the system is positive~\cite{Fujii:2003pa}. We shall explicitly demonstrate how negative $\zeta_i$ is allowed in the end of this section. 

It should be emphasized that $W_{\mu}$ does not couple to fermions directly. This is because there is no factor $i$ in the covariant derivative with $W_\mu$. As a result, $W_\mu$-dependent terms will cancel in the parentheses. Scalars can also couple to fermions with Yukawa interactions, which can lead to the generation of fermion mass, decay of scalars and reheating after inflation. 

At first sight, it seems there are many free parameters in Eq.~\ref{eq:act}. Actually, not all of them are independent. For example, if $\alpha \beta \neq 0 $, we can always rescale $\phi$ and $\varphi$ to make $|\alpha|= 1 = |\beta|$. Or if $|\zeta_i|\neq 0$, we can keep $\alpha$ and $\beta$ general but make $|\zeta_i|=1$. As long as one of $\zeta_i$ is not zero, we can always relabel the fields and rewrite the Lagrangian as following
\begin{align}\label{eq:act}
	\frac{\mathcal{L}}{\sqrt{-{g}}}= & \frac{\alpha}{2} \left(\varphi^{2}R-6\partial_{\mu}\varphi\partial^{\mu}\varphi\right) + \frac{\beta}{2} \left(\phi^{2}R-6\partial_{\mu}\phi\partial^{\mu}\phi\right) + \frac{1}{2}D_{\mu}\varphi D^{\mu}\varphi + 
	\frac{\zeta}{2}D_{\mu}\phi D^{\mu}\phi \nonumber \\
	& + \frac{i}{2}\left(\overline{\psi}\gamma^\mu D_\mu \psi -\overline{D_\mu \psi}\gamma^\mu \psi \right) 
	+ y\, \varphi \overline{\psi}\psi  + f \,\phi \overline{\psi}\psi - V-\frac{1}{4g_W^2}F_{\mu\nu}F^{\mu\nu},
\end{align}
where $\zeta$ can be positive or negative.

In additon to the general covariance of coordinate transformation, the above Lagrangian is invariant under the following local Weyl or scaling transformation
\begin{align}
& g_{\mu\nu}\left(x\right) \rightarrow {g}'_{\mu\nu}\left(x\right)=\lambda^{2}\left(x\right)g_{\mu\nu}\left(x\right),\;
\varphi\left(x\right)  \rightarrow {\varphi}'\left(x\right)=\lambda^{-1}\left(x\right)\varphi\left(x\right),
\nonumber \\
& \phi\left(x\right)  \rightarrow {\phi}'\left(x\right)=\lambda^{-1}\left(x\right)\phi\left(x\right),\;
\psi\left(x\right) \rightarrow {\psi}'\left(x\right)=\lambda^{-3/2}\left(x\right)\psi\left(x\right),\nonumber\\
& W_{\mu}\left(x\right)  \rightarrow {W}'_{\mu}\left(x\right)=W_{\mu}\left(x\right)-\partial_{\mu}\ln \lambda(x),
\end{align}
where the scale factor $\lambda\left(x\right)$ acts as a gauge parameter that may be taken in the domain $\lambda>0$. After fixing $\phi^{2}= v^2$, Einstein-Hilbert term $R$ can be recovered.  Weyl boson $W_{\mu}$ gets a mass due to the kinetic term of $\phi$ and fermion $\psi$ gets a mass from Yukawa interaction. Afterwards, the theory describes Einstein's gravity with a {\it non-minimally} coupled scalar $\varphi$, a massive gauge boson $W_{\mu}$ and a fermion $\psi$. We shall show that $\varphi$ can be responsible for cosmic inflation and $W_{\mu}$ can be a DM candidate. 

We understand that Weyl symmetry is broken by quantum corrections, namely the fields and parameters in the theory will be running and depend on the energy scale at which our physics is considered. Therefore, it can not be an exact symmetry. However, it is still useful to utilize Weyl symmetry at classical level since it provides a guiding principle for the starting Lagrangian, as we showed above. Also, if the couplings are small or the considered energy scale does not change much, we may neglect the running and treat Weyl symmetry as approximate.   

To demonstrate the above framework can provide a viable mechanism for cosmic inflation and DM, in the following we shall illustrate with a concrete example by fixing 
\begin{equation}
\beta=0 \textrm{ and } V=c(\varphi^2-\xi \phi^2)^2,
\end{equation} 
where $\xi>0$ is a numeric number in the Higgs-like potential. Since the rescaling of $\phi$ would rescale $\zeta$ and $\xi$ correspondingly, only the ratio $\zeta/\xi$ is physical. Hence, we can work in the basis that $\xi \equiv 1$ while keeping $\zeta$ free. However, it should be kept in mind that $\zeta$ in the rest of the paper can be effectively interpreted as $\zeta/\xi$.  We also emphasize that the above choice by no means is the only viable set, it is just a simple option that can elucidate the main physics. The general analysis for other options with $\beta\neq 0$ and different $V$s are explored in~\cite{Tang:2019olx}. 

After some algebra to make the kinetic terms canonical and to reorganize the resulting Lagrangian (see Appendix for the detailed derivation), we have
\begin{align}\label{eq:efflag}
\frac{\mathcal{L}}{\sqrt{-\overline{g}}}\supset & \frac{1}{2}\bar{R} + \frac{1}{2}\partial_\mu S \partial ^\mu S - \frac{c}{\alpha^2}\left[1-\frac{1}{\alpha \varphi^2(S)}\right]^{2}  +i\overline{\Psi}\gamma^{\mu}\partial_{\mu}\Psi -\frac{fv+y\varphi(S)}{\sqrt{\alpha}\varphi(S)}\overline{\Psi}\Psi \nonumber \\
&-\frac{1}{4g^2_W}F_{\mu\nu}F^{\mu\nu}+\frac{\zeta v^{2}+\varphi^{2}}{2\alpha\varphi^2(S)}\overline{W}_\mu\overline{W}^\mu,
\end{align}
where $v^2\equiv 1/\alpha$. Note that the new fields are related with the old ones through
\begin{align}
\bar{g}_{\mu\nu} = \lambda^{2}g_{\mu\nu},\; \lambda^2=\alpha \varphi^2,
\Psi =\lambda^{-3/2}\psi, \overline{W}_{\mu} = W_\mu -\partial _\mu \ln \sqrt{|\zeta/\alpha +\varphi^{2}|},
\end{align}
and the inflation field $S$ with canonical kinetic term is a function of $\varphi$,
\begin{equation}\label{eq:Sphi}
S = \frac{1}{\sqrt{\alpha}}\times\begin{cases}
\ln \dfrac{X}{1+\sqrt{1+X^2}}, & \zeta>0, X\equiv \dfrac{\varphi }{\sqrt{+\zeta/\alpha}},\\
\ln \dfrac{X}{1+\sqrt{1-X^2}}, & \zeta<0, X\equiv \dfrac{\varphi }{\sqrt{-\zeta/\alpha}}.
\end{cases}
\end{equation}
Or inversely $\varphi$ can be expressed as a function of $S$,
\begin{equation}\label{eq:phiS}
	\varphi(S)  = \begin{cases}
		\sqrt{+\zeta/\alpha}\dfrac{2Y}{1+Y^2}, & \zeta>0,\\
	 	\sqrt{-\zeta/\alpha}\dfrac{2Y}{1-Y^2}, & \zeta<0,
	\end{cases}
\end{equation}
where $Y=\exp{\sqrt{\alpha}S}$. In the vicinity of $X \ll 1$, we have simple relations, $S = (\ln X)/\sqrt{\alpha }$ or $\varphi \propto \exp{\sqrt{\alpha}S}$. From the above formalism, we can also see that both positive and negative $\zeta$ could give consistent theories, without theoretical pathology. However, if $\zeta = 0$, the kinetic term for $S$ would vanish and $S$ can be solved by equation of motion. This is because $S$ is defined by the differential equation to have canonical kinetic term (see Appendix for details),
\begin{equation}
	\frac{dS}{d\varphi}= \sqrt{\frac{\zeta v^2}{\alpha \varphi ^2 \left(\zeta v^2 + \varphi ^2\right)}}.
\end{equation}
The mass of $S$ can be obtained at the minimum $\varphi^2(S_0)=1/\alpha$, 
\begin{equation}
m^2_s=\frac{\partial^2 \mathcal{V}}{\partial S^2}= \dfrac{8c}{\alpha}\dfrac{1+\zeta}{\zeta}.
\end{equation}
Similarly, the mass of $\Psi$ is given by  $m_{\Psi}= (f+y)/\sqrt{\alpha}$ and the mass of $\overline{W}_\mu$ is calculated as $m_{\overline{W}}=g_W(\zeta + 1)/\alpha$, all in Planck unit. 

\section{Inflation}\label{sec:inf}
In this section, we elucidate how $S$ can be responsible for a successful inflation and contrast the predictions with experimental constraints. The potential of $S$ is given by 
\begin{equation}
\mathcal{V}\left(S\right)=\frac{c}{\alpha^{2}}\left[1-\frac{1}{\alpha \varphi^2(S)}\right]^{2},
\end{equation}
where $\varphi(S)$ is given in Eq.~\ref{eq:phiS}. The potential is very flat when $\varphi^2(S)\gg 1/\alpha$ where inflation happens, and its minimum $\mathcal{V}=0$ is reached at $\varphi^2(S)=1/\alpha$. 

\subsection{Inflationary Observables}
To compare with the observations, we calculate the standard slow-roll parameters~\cite{Liddle:1992wi},
\begin{align}\label{eq:slowroll}
	\epsilon & \equiv \frac{1}{2}\left(\frac{\mathcal{V}'}{\mathcal{V}}\right)^2=\frac{8 \alpha  \left(\zeta +\alpha  \varphi ^2\right)}{\zeta  \left(\alpha  \varphi^2-1\right)^2}, \\
	\eta &\equiv \frac{\mathcal{V}''}{\mathcal{V}}= -\frac{4 \alpha  \left[-4 \zeta +\alpha ^2 \varphi^4+\alpha  (2 \zeta -3) \varphi^2\right]}{\zeta  \left(\alpha  \varphi^2-1\right)^2},
\end{align}
where $'$ is denoted to the derivative over $S$. The slow-roll parameters are related with the cosmological observables, spectral index $n_s=1-6\epsilon+2\eta$ and tensor-to-scalar ratio $r=16\epsilon$, 
\begin{align}
n_s & = 1-\frac{8 \alpha  \left[2 \zeta +\alpha ^2 \varphi^4+\alpha  (2 \zeta +3) \varphi^2\right]}{\zeta  \left(\alpha  \varphi^2-1\right)^2},\\
r & = \frac{128 \alpha  \left(\zeta +\alpha  \varphi ^2\right)}{\zeta  \left(\alpha  \varphi^2-1\right)^2}.
\end{align} 
The e-folding number $N$ is defined as 
\begin{equation}\label{eq:efold}
N\equiv \ln \frac{a_{e}}{a_{i}}\simeq \int^{t_\textrm{end}}_t \mathcal{H}dt \simeq \int ^{S_i}_{S_e}\frac{dS}{\sqrt{2\epsilon}}= \int ^{\varphi_i}_{\varphi_e}\frac{dS}{d\varphi}\frac{d\varphi}{\sqrt{2\epsilon}}, 
\end{equation}
where $a_i (a_e)$ is the scale factor at initial (end) time of the inflation, $\varphi_i (\varphi_e)$ is the corresponding field value, and $\mathcal{H}$ is the Hubble parameter. Here $\varphi_e$ is determined by the violation of slow-roll condition, $\epsilon\sim 1$ or $\eta\sim 1$. To solve the flatness and horizon problems, the universe should inflate at least by $e^N$ with the typical $N\simeq 50\sim 60$ before inflation ends.
 
\begin{figure}
	\includegraphics[width=0.49\textwidth,height=0.45\textwidth]{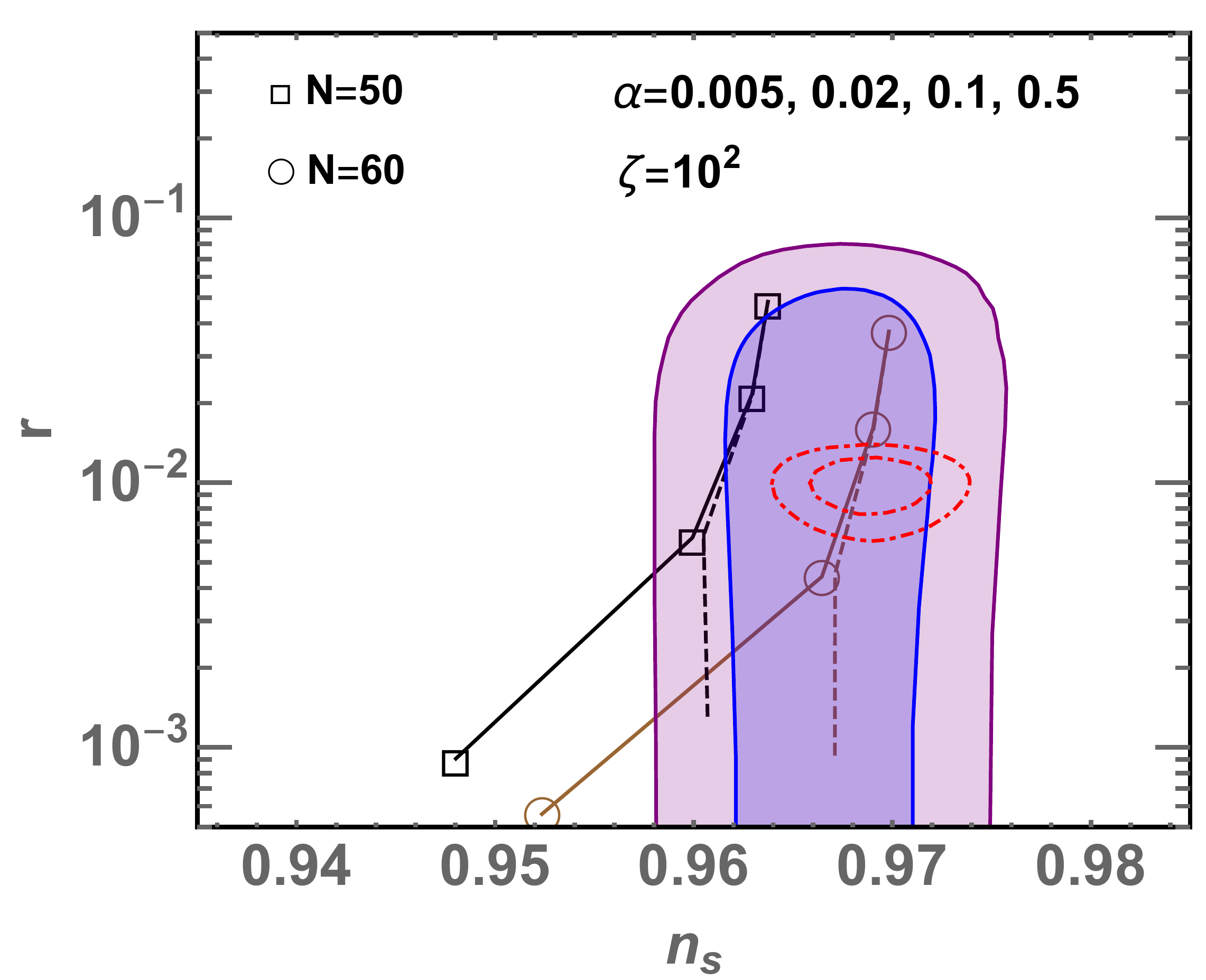}
	\includegraphics[width=0.49\textwidth,height=0.45\textwidth]{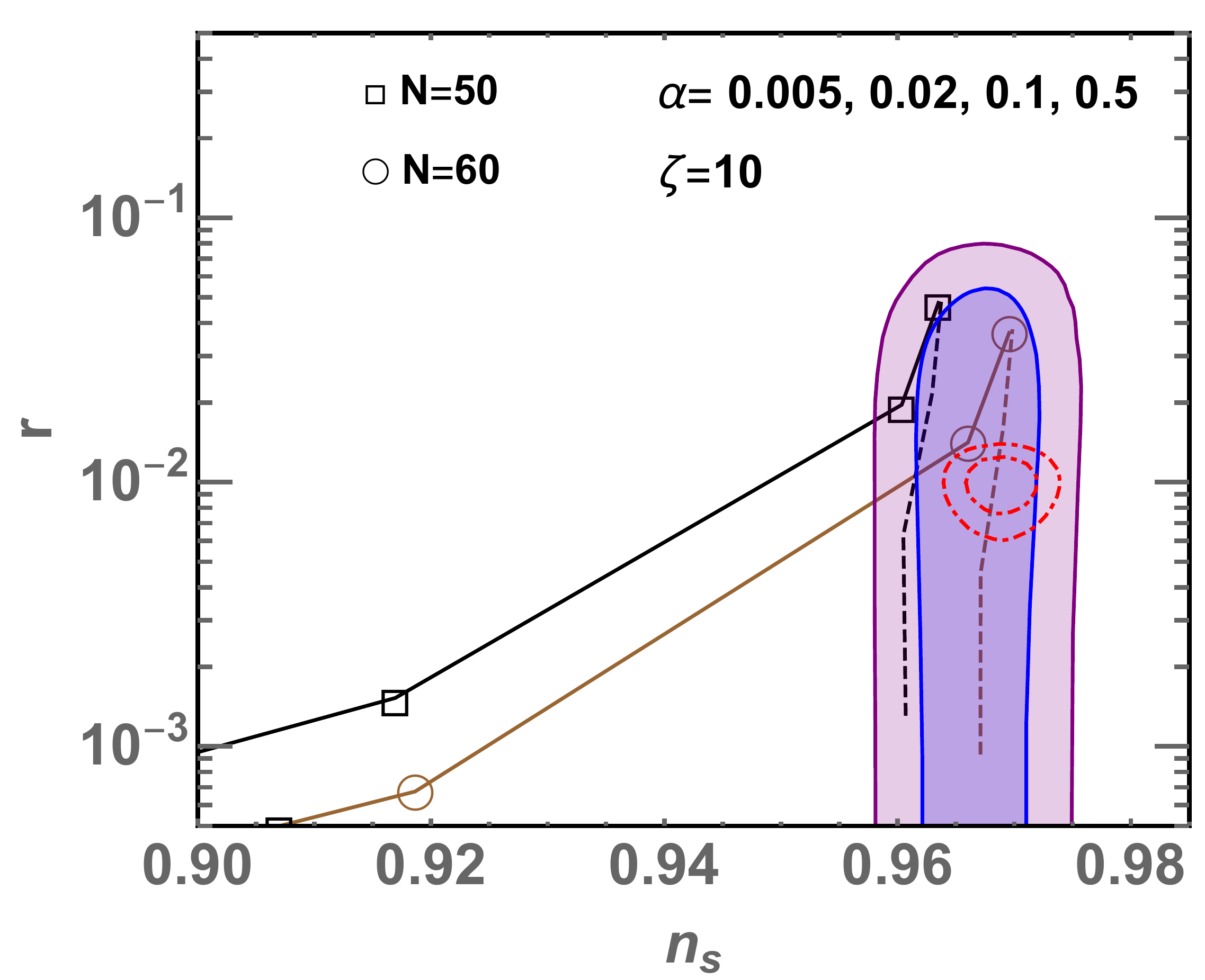}
	\caption{Illustration of $(n_s, r)$ when $\zeta=10^2$ (left panel) and $\zeta=10$ (right panel). The theoretical values of $(n_s, r)$ are shown for e-folding number $N=50$ (squares) and $60$ (circles) with $\alpha=0.005,0.02,0.1,0.5$ (from top to bottom), in comparison with the shaded regions allowed by {\it Planck}~\cite{Planck:2018} with 1-$\sigma$ and 2-$\sigma$ contours, and the future projection of CMB-S4~\cite{Abazajian:2016yjj} in smaller red contours. Along the solid lines, dashed lines indicate the cases with $\sqrt{|\zeta|/\alpha}\gg 30$.
		\label{fig:r-ns}}
\end{figure} 

In Fig.~\ref{fig:r-ns} we numerically solve the inflationary dynamics and present the calculated values of $(n_s, r)$ for e-folding number $N=50$ and $60$, in comparison with the allowed regions by {\it Planck}~\cite{Planck:2018}. We illustrate with ${\alpha}=0.005,0.02,0.1,0.5$ and $\zeta=10,10^2$. The projected sensitivities of the next-generation CMB experiments~\cite{Abazajian:2016yjj} are plotted as smaller red contours. Along the solid lines, the dashed lines represent the cases when $\sqrt{|\zeta|/\alpha}\gg 30$, an attractor behavior as $|\zeta|$ increasing. We have also checked that the predictions do not change for negative $\zeta$, as long as $\sqrt{|\zeta|/\alpha}\gg 30$. Below we give an intuitive explanation for this attractor behavior.

When $\sqrt{|\zeta|/\alpha}\gg 30$, analytic treatments are possible for qualitative understanding.  In such a case, in the field range that is relevant for the observable universe we have $\varphi \propto \exp (\sqrt{\alpha}S)$, a result of $X\ll 1$ in Eq.~\ref{eq:Sphi}. When $\alpha \gtrsim 0.1$, we find it is a good approximation in our model with analytical formula,  
\begin{equation}\label{eq:nsr}
n_s\simeq 1-\frac{2}{N},\; r\simeq \frac{2}{\alpha N^2},
\end{equation}
which are independent on $\zeta$, the so-called attractor behavior. This situation is very similar for the inflation in the induced gravity~\cite{Zee:1978wi} with the following Lagrangian,
\begin{equation}\label{eq:inducegravity}
\frac{\mathcal{L}}{\sqrt{-g}}=\frac{\alpha}{2}\varphi^2R + \frac{1}{2}\partial_{\mu} \varphi \partial^\mu \varphi - c\left(\varphi^2 - v^2\right)^2,
\end{equation}
where we have similar $(n_s, r)$ but with $\alpha$ in Eq.~\ref{eq:nsr} replaced with $\bar{\alpha}\equiv \dfrac{\alpha}{1+6\alpha}$. They can also be compared with $n_s\simeq 1-2/N$ and $r\simeq 12/ N^2$ in Starobinsky's inflation.

The measured overall amplitude of scalar power spectrum by {\it Planck}~\cite{Planck:2018}
\begin{equation}\label{eq:fluct}
\Delta_{s}^{2}\left(k\right)\approx \frac{1}{24\pi^{2}}\frac{\mathcal{V}}{\epsilon}\sim 2.2\times 10^{-9},
\end{equation}
requires $c \simeq 2\alpha\times 10^{-11}$ when $\sqrt{|\zeta|/\alpha}\gg 30$, a typical value in large-field inflation models. The Hubble parameter during inflation in this case can be estimated as $\mathcal{H}\sim \mathcal{V}^{1/2}/3\sim 1/\sqrt{\alpha} \times 10^{13}\GeV$. Thus, in this framework, only $\alpha$ and $\zeta$ are effectively free parameters.

\subsection{Reheating}

After inflation, inflaton field $S$ will oscillate around the potential minimum, transfer its energy to other fields and reheat the universe. The details of reheating depends on how inflaton and other fields are coupled. In the minimal model we considered, through the Yukawa interaction inflaton can decay into $\Psi$-pair, namely $S\rightarrow \overline{\Psi}\Psi$. When $S$ is much heavier than $\psi$, the decay width is 
\begin{equation}\label{eq:decayf}
\Gamma_S \sim m_S f^2/8\pi\times \left({\frac{\zeta-1}{\zeta}}\right)^2 .
\end{equation}
The reheating temperature $T_R$ is given by
\begin{equation}
T_R\simeq \sqrt{\Gamma_S M_p}\simeq 1.5f\frac{\zeta-1}{\zeta}\times 10^{16}\GeV
\end{equation}
This estimation shows that the reheating temperature can be as high as $T_R\sim 10^{15}\GeV$ for $f\sim 0.1$ and $\zeta \sim 2$. 

Note that the reheating temperature $T_R$ is referred to $\Psi$ only and can be different from the highest temperature of SM particles, because just after reheating $\Psi$ may not be in thermal equilibrium with SM. To connect $\Psi$ with SM, we can introduce a new $U(1)$ gauge symmetry with coupling $g$ and gauge boson $V_\mu$, under which both $\Psi$ and SM fermions are charged. These new interactions still respect the local Weyl symmetry and do not affect our previous discussions. We find that depending the interaction strength, $\Psi$ would reach thermal equilibrium with SM through the scattering and annihilation processes mediated by $V_\mu$, at temperature $T_h\simeq g^4M_p\left(\leq T_R\right)$ when the scattering rate $\Gamma_\Psi = n_\Psi \sigma_\Psi \sim g^4 T_h$ is equal to the Hubble parameter $\mathcal{H}\sim T_h^2/M_p$. By changing the interaction strength $g$, we can get different $T_h$ for SM particles. For instance, we can have $T_h\sim 10^{15}\GeV$ for $g\sim 0.14$ and $T_h\sim 10^{5}\GeV$ for $g\sim 5\times 10^{-4}$. 

\section{Weyl Boson as Dark Matter}\label{sec:dark}
It is apparent that in Eq.~(\ref{eq:efflag}) there is a discrete $Z_2$ symmetry for Weyl gauge boson $\overline{W}_{\mu}$, $\overline{W}_{\mu}\rightarrow -\overline{W}_{\mu}$. Then it would be tempting to ask whether $\overline{W}_{\mu}$ can be a DM candidate. This was first pointed out in Refs.~\cite{Cheng:1988zx, Cheng:2004sd} in a different context and investigated further~\cite{Wei:2006gv, Kashyap:2012kg}. However, the claims in the literature were controversial. The author in Refs.~\cite{Cheng:1988zx, Cheng:2004sd} stated that Weyl gauge boson was stable through an illustation with Higgs boson and sigma model, while later it was shown to be decaying when there are two scalars \cite{Kashyap:2012kg}. Below we shall set down the issue by presenting a general proof that $\overline{W}_{\mu}$ is stable, regardless of how many scalars are present. 

\subsection{Proof of Stability}
We consider the case with $N$ scalars whose Lagrangian is given by
\begin{equation}
\frac{\mathcal{L}}{\sqrt{-g}}\supset \sum_{i=1}^{N}\frac{\alpha_{i}}{2}\left(\phi_{i}^{2}R-6\partial_{\mu}\phi_{i}\partial^{\mu}\phi_{i}\right)+\frac{1}{2}\sum_{i=1}^{N}\zeta_{i}D_{\mu}\phi_{i}D^{\mu}\phi_{i}-V\left(\phi_{i}\right)-\frac{1}{4g_{W}^{2}}F_{\mu\nu}F^{\mu\nu}.
\end{equation}
As general as possible, we have included the factor $\zeta_i$ in the front of the covariant kinetic term, $\mathcal{L}_{k}\equiv \frac{1}{2}\sum_{i=1}^{N}\zeta_{i}D_{\mu}\phi_{i}D^{\mu}\phi_{i}$. As shown in previous sections, $\zeta_i$ is not necessarily positive. To show the stability of Weyl gauge boson, we  rewrite the covariant kinetic term as
\begin{align}\label{eq:proof}
2\mathcal{L}_{k} &=\sum_{i=1}^{N}\left[\zeta_{i}\partial_{\mu}\phi_{i}\partial^{\mu}\phi_{i}-W_{\mu}\partial^{\mu}\left(\zeta_{i}\phi_{i}^{2}\right)+\zeta_{i}\phi_{i}^{2}W_{\mu}W^{\mu}\right] \nonumber \\
&=\Xi\left(W_{\mu}W^{\mu}-W_{\mu}\frac{\partial^{\mu}\Xi}{\Xi}\right)+\sum_{i=1}^{N}\zeta_{i}\partial_{\mu}\phi_{i}\partial^{\mu}\phi_{i}\nonumber \\
&=\Xi\left(W_{\mu}-\frac{1}{2}\partial^{\mu}\ln\Xi\right)^{2}-\frac{1}{4}\Xi\times\left(\partial^{\mu}\ln\Xi\right)^{2}+\sum_{i=1}^{N}\zeta_{i}\partial_{\mu}\phi_{i}\partial^{\mu}\phi_{i} \nonumber \\
&=\Xi\overline{W}_{\mu}\overline{W}^{\mu}-\frac{\Xi}{4}\left[\partial_{\mu}\ln\Xi\right]^{2}+\sum_{i=1}^{N}\zeta_{i}\partial_{\mu}\phi_{i}\partial^{\mu}\phi_{i},
\end{align}
where we have defined $
\Xi\equiv\sum_{i=1}^{N}\zeta_{i}\phi_{i}^{2},\;\overline{W}_{\mu} = W_{\mu}-\dfrac{1}{2}\partial^{\mu}\ln\Xi.$ The above derivation does not depend on the potential form and is also valid for Higgs-like potential. One may wonder whether the proof still holds if there are other scalars that were not included in $\Xi$ from the beginning, like standard model Higgs or hidden scalar. In the Appendix, we show the proof is still valid in the presence of additional scalars that coupled to $W_\mu$ covariantly. 

Note that the redefinition of Weyl gauge field does not affect $F_{\mu\nu}$ due to its anti-symmetric identity.
Since $\overline{W}_\mu$ can get mass and interact only through the $\Xi\overline{W}_{\mu}\overline{W}^{\mu}$ term, it is now clear that $Z_{2}$ symmetry for $\overline{W}_{\mu}$ is manifest, even if taking the radiative correction into account. As a DM candidate, it would be stable.

The rest procedures go as standard. Define $\Omega^{2}\equiv \sum_{i=1}^{N}\alpha_{i}\phi_{i}^{2}$, we are now left with
\begin{equation}
\frac{\mathcal{L}}{\sqrt{-g}}\supset \frac{1}{2}\Omega^{2}R+\frac{1}{2}\sum_{i}^{N}\left(\zeta_{i}-6\alpha_{i}\right)\partial_{\mu}\phi_{i}\partial^{\mu}\phi_{i}-\frac{\left(\partial_{\mu}\Xi\right)^{2}}{8\Xi}-V\left(\phi_{i}\right)-\frac{1}{4g_{W}^{2}}F_{\mu\nu}F^{\mu\nu}+\frac{1}{2}\Xi\overline{W}_{\mu}\overline{W}^{\mu}.
\end{equation}
To make things more familar, we can make conformal transformation $\overline{g}_{\mu\nu}=\Omega^{2}g_{\mu\nu}$ and change to Einstein frame. Then we obtain
\begin{align}\label{eq:DM}
\frac{\mathcal{L}}{\sqrt{-\bar{g}}}=&  \frac{1}{2}\overline{R}+\frac{1}{2\Omega^{2}}\left[6\Omega^2\partial_{\mu}\ln\Omega\partial^{\mu}\ln\Omega+ \sum_{i}^{N}\left(\zeta_{i}-6\alpha_{i}\right)\partial_{\mu}\phi_{i}\partial^{\mu}\phi_{i}-\frac{\left(\partial_{\mu}\Xi\right)^{2}}{4\Xi}\right] -\frac{V\left(\phi_{i}\right)}{\Omega^{4}}\nonumber \\
&-\frac{1}{4g_{W}^{2}}F_{\mu\nu}F^{\mu\nu}+\frac{1}{2\Omega^{2}}\Xi\overline{W}_{\mu}\overline{W}^{\mu},
\end{align}
where we have used the following relation,
\begin{equation}
R=\Omega^{2}\left[\bar{R}+6\bar{g}^{\mu\nu}\partial_{\mu}\ln\Omega\partial_{\nu}\ln\Omega\right]. 
\end{equation}
The non-canonical kinetic term for $\phi_i$ in Eq.~\ref{eq:DM} is rather complicated. Only in several special cases there are analytic and transparent reductions, see our thorough analysis in Ref.~\cite{Tang:2019olx} for details. However, for the illustration of $\overline{W}_\mu$ as a DM candidate, it is now sufficient to use the Lagrangian in Sec.~\ref{sec:frame}, Eq.~\ref{eq:efflag}. 

\subsection{Relic Density}
Admittedly, for $g_W \sim 1$, $m_W$ would be naturally around Planck scale, which is too heavy to be produced in the early universe and whose cosmological consequence is uncertain in this context. However, if we temporarily put aesthetic reasons aside, and treat $g_W$ as a free parameter, tiny $g_W$ would induce a light $\overline{W}_\mu$ that can be produced abundantly, a potential DM candidate with $Z_2$ symmetry.

To demonstrate in principle there are parameter spaces that can give rise to the correct relic abundance for $\overline{W}_\mu$, we focus on the interactions in Eq.~\ref{eq:efflag}. The interaction between inflaton $S$ and Weyl boson $\overline{W}_\mu$ can be obtained by expanding $\varphi$ around the potential minimum, $\varphi^2(S_0)=1/\alpha$. For $\zeta>0$, in the linear order we have 
\begin{equation}\label{eq:intaction}
\frac{1}{\varphi}\simeq \sqrt{\alpha }+ \sqrt{\frac{\alpha}{\zeta}}\sinh (\sqrt{\alpha}S_0)\sqrt{\alpha}s = 
\sqrt{\alpha } + \alpha \frac{\zeta -1}{\zeta}s,
\end{equation}
where $s\equiv S-S_0, \sinh (\sqrt{\alpha}S_0)=\sqrt{\zeta -1}$ and $\cosh (\sqrt{\alpha}S_0)=\sqrt{\zeta}$. Then, we can obtain the linear interaction term
\begin{equation}
-\frac{\sqrt{\zeta (\zeta -1)}}{\alpha}s\overline{W}_{\mu}\overline{W}^{\mu}.
\end{equation}
Note that explicitly each $\overline{W}_{\mu}$ has a $g_W$ factor in it. 

As mentioned above, if $g_W\sim 1$, we would expect $m_W\sim M_P$ and it is different to produce such heavy particle in the early universe. If $g_W\ll 1$, the interaction would be too weak to keep it in thermal equilibrium. Therefore, $\overline{W}_\mu$ can not be a thermal DM. Nevertheless, we may consider non-thermal production. Below, we discuss two possible mechanisms.
 
In the case that $g_W$ is extremely small, we may neglect the above interaction and only consider the gravitational production~\cite{Graham:2015rva, Ema:2019yrd} which gives the relic abundance $\Omega_W$, 
\begin{equation}
\Omega_W \simeq \Omega_{\textrm{DM}}\times \sqrt{\frac{m_W}{6\times 10^{-11}\GeV}}\times \left(\frac{\mathcal{H}}{10^{13}\GeV}\right)^2,
\end{equation}
where $\Omega_{\textrm{DM}}\simeq 0.25$. In this case, $g_W\sim 10^{-29}$, a very tiny coupling, which indicates how challenging it is to detect such a DM particle.

When $g_W$ can not be neglected, the above production mechanism would not apply. We may consider an alternative production from inflaton's decay. We can calculate the decay width $s\rightarrow \overline{W}_\mu + \overline{W}_\mu$,
\begin{equation}\label{eq:decayW}
\Gamma\left(s\rightarrow \overline{W}_\mu + \overline{W}_\mu\right)=\frac{g_{W}^{4}\zeta (\zeta-1)}{32\pi\alpha^2}\frac{M_{P}^{2}m_{s}^{3}}{m_{W}^{4}}\sqrt{1-x_{W}}\left(1-x_{W}+\frac{3}{4}x_{W}^{2}\right),
\end{equation}
where $x_{W}=4m_{W}^{2}/m_{s}^{2}$. We denote $\mathcal{B}r$ as the branch ratio of the above decay mode, which can be estimated as $\propto m^2_s/(f^2M^2_P)$ by taking the ratio of Eq.~\ref{eq:decayW} to Eq.~\ref{eq:decayf}. The relic abundance from inflaton decay is evaluated as
\begin{equation}
	\frac{\rho_W}{s}\sim \frac{2m_Wn_s\mathcal{B}r}{s}\sim \frac{2m_WT^4_R\mathcal{B}r}{m_sT^3_R}=\frac{2m_WT_R\mathcal{B}r}{m_s},
\end{equation}
where $\rho_W$ is energy density of $\overline{W}_\mu$ and $s$ is the entropy density.  Putting in the relevant quantities, we can actually simplify the above formula to
\begin{equation}
\frac{\rho_W}{s}= \frac{2m_WT_R\mathcal{B}r}{m_s}\simeq \frac{2m_W}{f}\left(\frac{m_s}{m_P}\right)^{{3}/{2}}\frac{\alpha^2\zeta^2}{(\zeta + 1)^4}\simeq 10^{-9}\GeV\times \left(\frac{m_W}{\TeV}\right)\left(\frac{0.1}{f}\right),
\end{equation}
where in the last step we have used $\zeta \sim 10^2$ and $\alpha \sim 0.1$ for consistent inflation.
For $\dfrac{\rho_W}{s}\sim 10^{-9}\GeV$, we can have the correct relic abundance of DM. If we restrict $f\lesssim 4\pi$ for perturbativity, we would have an upper bound, $m_W\lesssim 100\TeV$.

\section{Conclusion}\label{sec:concl}
We have presented a theoretical study that the original Weyl scaling symmetry can provide a unified framework to explain the cosmic inflation and DM simultaneously. The inspired inflationary scenario has a Weyl-symmetric Lagrangian from the beginning. After the generation of Planck scale, the potential can be flat enough to allow a slow-roll inflation. The theoretical values of scalar spectral index and tensor-to-ratio are well consistent with current observations and can be tested in future CMB experiments, which can be clearly seen in Fig.~(\ref{fig:r-ns}). 

We have also clarified and proved the stability of Weyl gauge boson and demonstrated it can be a DM candidate if the gauge coupling is tiny, thanks to the $Z_2$ symmetry. The stability is valid for any theory with multiple scalars, as long as they are coupled to Weyl gauge boson covariantly. The mass of Weyl boson is generally very heavy unless the gauge coupling is very tiny, which then requires non-thermal productions. We discussed two viable mechanisms, gravitational production and inflaton's decay. However, detection of such DM would be challenging since its couplings to standard model particles are very small. 

\section*{Acknowledgments}

YT is partly supported by Natural Science Foundation of China (NSFC) under Grants No.~11851302 and the Grant-in-Aid for Innovative Areas No.16H06490. YT is grateful to Takeo Moroi for enlightening discussions. YLW is supported in part by NSFC under Grants No.~11851302, No.~11851303, No.~11690022, No.~11747601, and the Strategic Priority Research Program of the Chinese Academy of Sciences under Grant No. XDB23030100 as well as the CAS Center for Excellence in Particle Physics (CCEPP).

\section*{Appendix}
\subsection{Derivation of Eq.~\ref{eq:efflag}}
Here, we give the detailed derivation of Eq.~\ref{eq:efflag} in the main context. We start with the Lagrangian for two real scalars ( $\varphi$ and $\phi$
) and a fermion $\psi$,
\begin{align*}
\mathcal{L}\supset \sqrt{-g}\Big[& \frac{\alpha}{2} \left(\varphi^{2}R-6\partial_{\mu}\varphi\partial^{\mu}\varphi\right) + \frac{\beta}{2} \left(\phi^{2}R-6\partial_{\mu}\phi\partial^{\mu}\phi\right) + \frac{\zeta_1}{2}D_{\mu}\varphi D^{\mu}\varphi + 
\frac{\zeta_2}{2}D_{\mu}\phi D^{\mu}\phi \nonumber \\
& + \frac{i}{2}\left(\overline{\psi}\gamma^\mu D_\mu \psi -\overline{D_\mu \psi}\gamma^\mu \psi \right) 
+ y\, \varphi \overline{\psi}\psi  + f \,\phi \overline{\psi}\psi -V(\phi, \varphi)-\frac{1}{4g_W^2}F_{\mu\nu}F^{\mu\nu}\Big],
\end{align*}
where $D_{\mu}=\partial_{\mu}-W_{\mu},\;\gamma^{\mu}D_{\mu}\equiv\gamma^{a}\chi_{a}^{\mu}D_{\mu},\;\chi_{a}^{\mu}\chi_{b}^{\nu}\eta^{ab}=g^{\mu\nu}.$
As explained in the main text, for an illustration, we fix the following model parameters
\[
\beta=0,\,\zeta_{1}=1,\,\zeta_{2}\equiv\zeta\textrm{ and }V=c(\varphi^{2}-\phi^{2})^{2},
\]
and set $\phi^{2}=v^{2}\equiv 1/\alpha$, thanks to the freedom from the local Weyl gauge symmetry. Then we have the following Lagrangian in Jordan frame,
\begin{align*}
	\mathcal{L}\supset\sqrt{-g} & \left[ \frac{\alpha}{2}\left(\varphi^{2}R-6\partial_{\mu}\varphi\partial^{\mu}\varphi\right)+\frac{1}{2}D_{\mu}\varphi D^{\mu}\varphi-c(\varphi^{2}-v^{2})^{2}\right.\\
	& \left.+i\overline{\psi}\gamma^{\mu}\partial_{\mu}\psi-f\,v\overline{\psi}\psi-y\,\varphi\overline{\psi}\psi-\frac{1}{4g_W^2}F_{\mu\nu}F^{\mu\nu}+\frac{1}{2}\zeta v^{2}W_{\mu}W^{\mu}\right].
\end{align*}
Note that $\varphi$ is not minimally coupled to gravity. To compare with Einstein's gravity and observations, we can redefine the fields by conformal transformations,
\[
\overline{g}_{\mu\nu}=\lambda^{2}g_{\mu\nu},\;\Psi=\lambda^{-3/2}\psi, \; 
\lambda^{2}\equiv \alpha\varphi^{2},
\]
and rewrite the Lagrangian as
\begin{align*}
	\frac{\mathcal{L}}{\sqrt{-\overline{g}}}\supset & \frac{1}{2}\bar{R} +\frac{1}{2\lambda^2}\partial_\mu \varphi \partial ^\mu \varphi - \frac{c}{\lambda^4}(\varphi^{2}-v^{2})^{2}  +i\overline{\Psi}\gamma^{\mu}\partial_{\mu}\Psi-\lambda^{-1}\left(fv+y\varphi\right)\overline{\Psi}\Psi\\
	& -\lambda^{-2}\varphi\partial_{\mu}\varphi W^{\mu}+\frac{1}{2}\lambda^{-2}\left(\zeta v^{2}+\varphi^{2}\right)W_{\mu}W^{\mu}-\frac{1}{4g^2_W}F_{\mu\nu}F^{\mu\nu}.
\end{align*}
We can rearrange the gauge interactions
\begin{align*}
\frac{1}{2\lambda^{2}}\left[\left(\zeta v^{2}+\varphi^{2}\right)W_{\mu}W^{\mu} -W^{\mu}\partial_{\mu}\varphi^2 \right]= \frac{\zeta v^{2}+\varphi^{2}}{2\lambda^2}\overline{W}_\mu\overline{W}^\mu - \frac{\varphi^2 \partial_\mu \varphi \partial ^\mu \varphi }{2\lambda^2 \left(\zeta v^{2}+\varphi^{2}\right)},
\end{align*}
where the new Weyl field has a gauge transformation $\overline{W}_\mu = W_\mu - \frac{1}{2}\partial _\mu \ln |\zeta v^{2}+\varphi^{2}|$ and the last term in the above equation would contribute additionally to the kinetic term for $\varphi$, which in total is given by
\begin{align*}
\frac{1}{2\lambda^2}\left[1-\frac{\varphi^2  }{ \zeta v^{2}+\varphi^{2}}\right]\partial_\mu \varphi \partial ^\mu \varphi = \frac{1}{2}\frac{\zeta v^2}{\alpha \varphi ^2 \left(\zeta v^2 + \varphi ^2\right)}\partial_\mu \varphi \partial ^\mu \varphi\equiv \frac{1}{2}\partial_\mu S \partial^\mu S.
\end{align*}
Here we have defined the new field $S$ through
\begin{equation*}
\frac{dS}{d\varphi}= \sqrt{\frac{\zeta v^2}{\alpha \varphi ^2 \left(\zeta v^2 + \varphi ^2\right)}}.
\end{equation*}
One can immediately notice $\zeta\neq 0$, otherwise $\varphi$ is not a dynamical field. And $\zeta$ can be negative as long as $\left(\zeta v^2 + \varphi ^2\right)\zeta v^2>0$. 

Generally we have the solutions for $S=S(\varphi)$,
\begin{equation*}
S = \frac{1}{\sqrt{\alpha}}\times\begin{cases}
	\ln \dfrac{X}{1+\sqrt{1+X^2}}, & \zeta>0, X\equiv \dfrac{\varphi }{\sqrt{+\zeta/\alpha}},\\
	\ln \dfrac{X}{1+\sqrt{1-X^2}}, & \zeta<0, X\equiv \dfrac{\varphi }{\sqrt{-\zeta/\alpha}}.
\end{cases}
\end{equation*}
Or we can obtain inversely
\begin{equation*}
	\varphi  = \begin{cases}
	 \sqrt{+\zeta/\alpha}\dfrac{2Y}{1+Y^2}, & \zeta>0,\\
	 \sqrt{-\zeta/\alpha}\dfrac{2Y}{1-Y^2}, & \zeta<0,
	\end{cases}
\end{equation*}
where $Y=\exp{\sqrt{\alpha}S}$. In the vicinity of $X<<1$, we have $S = (\ln X)/\sqrt{\alpha }$ or $\varphi \propto \exp{\sqrt{\alpha}S}$. 
Finally, the Lagrangian can be rewritten as
\begin{align*}
	\frac{\mathcal{L}}{\sqrt{-\overline{g}}}\supset & \frac{1}{2}\bar{R} +\frac{1}{2}\partial_\mu S \partial ^\mu S - \frac{c}{\alpha^2}\left[1-\frac{v^{2}}{\varphi^2(S)}\right]^{2}  +i\overline{\Psi}\gamma^{\mu}\partial_{\mu}\Psi-\frac{fv+y\varphi(S)}{\sqrt{\alpha}\varphi(S)}\overline{\Psi}\Psi\\
	&-\frac{1}{4g^2_W}F_{\mu\nu}F^{\mu\nu}+\frac{\zeta v^{2}+\varphi^{2}}{2\alpha\varphi^2(S)}\overline{W}_\mu\overline{W}^\mu.
\end{align*}

\subsection{The Second Step for The Proof}
Let us assume there is another scalar field $\Phi$ that couples to $W_\mu$ covariantly ($\zeta D_\mu \Phi D^\mu \Phi$), but was not included in the definition of $\Xi$ in Eq.~\ref{eq:proof}, then we would have for the total kinetic term
\begin{equation}\label{eq:two-step}
	2\mathcal{L}_{k} 
	=\Xi\overline{W}_{\mu}\overline{W}^{\mu}-\frac{\Xi}{4}\left[\partial_{\mu}\ln\Xi\right]^{2}+\sum_{i=1}^{N}\zeta_{i}\partial_{\mu}\phi_{i}\partial^{\mu}\phi_{i} + \zeta\left[\partial_{\mu}\Phi\partial^{\mu}\Phi-W_{\mu}\partial^{\mu}\Phi^{2}+\Phi^{2}W_{\mu}W^{\mu}\right]. 
\end{equation}
We shall prove with the above Lagrangian can be rewritten as 
\begin{align}
2\mathcal{L}_{k}
=\widetilde{\Xi}\widetilde{W}_\mu \widetilde{W}^\mu -\frac{\widetilde{\Xi}}{4}\left[\partial_{\mu}\ln\widetilde{\Xi}\right]^{2} +\sum_{i=1}^{N+1}\zeta_{i}\partial_{\mu}\phi_{i}\partial^{\mu}\phi_{i},
\end{align}
where $\phi_{N+1}\equiv \Phi,\zeta_{N+1}\equiv \zeta $, $\widetilde{W}_\mu = \overline{W}_\mu - \frac{1}{2}\partial_{\mu}\ln \dfrac{\widetilde{\Xi}}{\Xi},\; \widetilde{\Xi}= \Xi +\zeta \Phi^2 = \sum_{i=1}^{N+1}\zeta_{i}\phi_i^2$.

Replace $W_{\mu} = \overline{W}_{\mu}+\dfrac{1}{2}\partial^{\mu}\ln\Xi$ in Eq.~\ref{eq:two-step} and combine $\overline{W}_{\mu}\overline{W}^{\mu}$ terms, we get 
\begin{align}
2\mathcal{L}_{k}
=&\left(\Xi+\zeta\Phi^2\right)\overline{W}_{\mu}\overline{W}^{\mu} - \overline{W}_\mu \partial^\mu \left(\zeta \Phi^2\right) -\frac{1}{2}\partial_{\mu}\ln\Xi\partial^{\mu}\left(\zeta\Phi^{2}\right)+\zeta\Phi^{2}\overline{W}_{\mu}\partial^{\mu}\ln\Xi \nonumber\\
&{}+\frac{1}{4}\zeta\Phi^{2}\left[\partial^{\mu}\ln\Xi\right]^2-\frac{\Xi}{4}\left[\partial_{\mu}\ln\Xi\right]^{2}+\sum_{i=1}^{N+1}\zeta_{i}\partial_{\mu}\phi_{i}\partial^{\mu}\phi_{i}.
\end{align}
Note that there is a mixing term $\overline{W}_\mu \partial^\mu \left(\zeta \Phi^2\right)$ which appears to induce the decay of $\overline{W}_\mu$. However, this term actually can be canceled by a gauge transformation of $\overline{W}_\mu$, as we shall show below.
\begin{align}
2\mathcal{L}_{k}
=&\left(\Xi+\zeta\Phi^{2}\right)\left\{ \overline{W}_{\mu}\overline{W}^{\mu}-\overline{W}_{\mu}\partial^{\mu}\ln\left(\frac{\Xi+\zeta\Phi^{2}}{\Xi}\right)+\frac{1}{4}\left[\partial_{\mu}\ln\left(\frac{\Xi+\zeta\Phi^{2}}{\Xi}\right)\right]^{2}\right\}  \nonumber\\
&-\frac{1}{4}\left(\Xi+\zeta\Phi^{2}\right)\left[\partial_{\mu}\ln\left(\frac{\Xi+\zeta\Phi^{2}}{\Xi}\right)\right]^{2}+\overline{W}_{\mu}\partial^{\mu}\left(\Xi+\zeta\Phi^2\right)-\left(\Xi+\zeta\Phi^{2}\right)\overline{W}_{\mu}\partial^{\mu}\ln\left(\Xi\right)  \nonumber\\
&- \overline{W}_\mu \partial^\mu \left(\zeta \Phi^2\right) -\frac{1}{2}\partial_{\mu}\ln\Xi\partial^{\mu}\left(\zeta\Phi^{2}\right)+\zeta\Phi^{2}\overline{W}_{\mu}\partial^{\mu}\ln\Xi +\frac{1}{4}\zeta\Phi^{2}\left[\partial^{\mu}\ln\Xi\right]^2\nonumber\\
&{}-\frac{\Xi}{4}\left[\partial_{\mu}\ln\Xi\right]^{2}+\sum_{i=1}^{N+1}\zeta_{i}\partial_{\mu}\phi_{i}\partial^{\mu}\phi_{i}.
\end{align}
We immediately realize that all the linear terms of $\overline{W}_\mu$ in the second and third lines cancel completely. So we have 
\begin{align}
2\mathcal{L}_{k}
=&\left(\Xi+\zeta\Phi^{2}\right)\left[ \overline{W}_{\mu}-\frac{1}{2}\partial_{\mu}\ln\left(\frac{\Xi+\zeta\Phi^{2}}{\Xi}\right)\right]^{2} +\sum_{i=1}^{N+1}\zeta_{i}\partial_{\mu}\phi_{i}\partial^{\mu}\phi_{i} + \mathcal{C}. \\
\mathcal{C}\equiv &-\frac{1}{4}\left(\Xi+\zeta\Phi^{2}\right)\left[\partial_{\mu}\ln\left(\Xi+\zeta\Phi^{2}\right)-\partial_{\mu}\ln {\Xi}\right]^{2}\nonumber \\
&-\frac{\Xi}{4}\left[\partial_{\mu}\ln\Xi\right]^{2} -\frac{1}{2}\partial_{\mu}\ln\Xi\partial^{\mu}\left(\zeta\Phi^{2}\right) +\frac{1}{4}\zeta\Phi^{2}\left[\partial^{\mu}\ln\Xi\right]^2.
\end{align}
Though tedious, it is however straightforward to show
\begin{equation}
\mathcal{C}=-\frac{1}{4}\left(\Xi+\zeta\Phi^{2}\right)\left[\partial_{\mu}\ln\left(\Xi+\zeta\Phi^{2}\right)\right]^{2}. 
\end{equation}
Eventually, we have obtained the kinetic term for $N+1$ scalars by two-step procedure, 
\begin{align}
2\mathcal{L}_{k}
=&\widetilde{\Xi}\widetilde{W}_\mu \widetilde{W}^\mu +\sum_{i=1}^{N+1}\zeta_{i}\partial_{\mu}\phi_{i}\partial^{\mu}\phi_{i} -\frac{\widetilde{\Xi}}{4}\left(\partial_{\mu}\ln\widetilde{\Xi}\right)^{2}, \\
\widetilde{W}_\mu = & {}\ \overline{W}_{\mu}-\frac{1}{2}\partial_{\mu}\ln\left(\frac{\Xi+\zeta\Phi^{2}}{\Xi}\right) = 
W_\mu - \frac{1}{2}\partial_{\mu}\ln \widetilde{\Xi},
\end{align}
where $\tilde{\Xi}= \Xi +\zeta \Phi^2 = \sum_{i=1}^{N+1}\zeta_{i}\phi_i^2$. Evidently, $Z_2$ symmetry for $\widetilde{W}$ is manifest. 



%

\end{document}